\documentclass[twocolumn,showpacs,preprintnumbers,amsmath,amssymb,prl]{revtex4}
\usepackage[dvipdfm]{graphicx}
\usepackage{dcolumn}
\usepackage{bm}
\begin{document}

\title{Prospects for Narrow-line Cooling of KRb Molecules \\ in
Rovibrational Ground State}

\author{J. Kobayashi}

\email{kobayashi@atomtrap.t.u-tokyo.ac.jp}
\author{K. Aikawa}
\altaffiliation{Present address: Universit\"{a}t Innsbruck, 6020 Innsbruck, Austria.}
\author{K. Oasa}
\altaffiliation{Present address: TOSHIBA Corporation, Tokyo 105-8001, Japan}
\author{S. Inouye}
\affiliation{School of Engineering, The University of Tokyo, Yayoi, Bunkyo-ku, Tokyo 113-8656, Japan}

\date{\today}
\begin{abstract}
We propose and experimentally investigate a scheme for narrow-line cooling of KRb molecules in the rovibrational ground state. We show that the spin-forbidden
$\mathrm{X^1\Sigma^+} \rightarrow \mathrm{b^3\Pi_{0^+}}$ transition of KRb is ideal for realizing narrow-line laser cooling of molecules because it has highly diagonal Franck-Condon factors and narrow linewidth. In order to confirm the prediction, we performed the optical and microwave spectroscopy of ultracold $^{41}$K$^{87}$Rb molecules, and determined the linewidth ($2\pi\times$ 4.9(4) kHz) and Franck-Condon factors for
the $\mathrm{X^1\Sigma^+} (v''=0) \rightarrow \mathrm{b^3\Pi_{0^+}} (v'=0)$ transition (0.9474(1)). This result opens the door towards all-optical production of polar molecules at sub-microkelvin temperatures.
\end{abstract}
\maketitle

Recently, significant progress has been made in the field of cold molecules. Innovative ideas are implemented one after another, making cold molecules an attractive new tool to test fundamental physics \cite{meer08,carr09,krem09}.
However, in spite of serious efforts by many groups, a quantum degenerate gas of polar molecules remains an elusive goal. The JILA group got closest to this goal by combining magneto-association with stimulated Raman adiabatic passage (STIRAP); they reached a phase-space density of the order of 0.1 \cite{ni08}. In that experiment, they were limited by the magneto-association efficiency, which was determined by the density overlap of the rubidium and potassium clouds \cite{cumb13}.

Motivated by this situation, we studied the possibility of laser cooling bi-alkali-metal molecules in the rovibrational ground state. Typically, laser cooling molecules is difficult because molecules have vibrational and rotational degrees of freedom.
Recently, however, several groups succeeded in laser cooling molecules by selecting special types
of molecules (such as SrF \cite{shum10}, YO \cite{humm13}, and CaF \cite{zhel03}) and constructing a quasicycling transition with minimal increase in laser complexity.
The solution to the problem was to find an optical transition with highly diagonal Franck-Condon factors \cite{rosa04}.
We observed such a transition in KRb --- one of the most popular species in the field of cold molecules.
We claim that the intercombination transition $\mathrm{X^1\Sigma^+}
 \rightarrow \mathrm{b^3\Pi_{0^+}} $ of KRb is ideal for realizing narrow-line cooling of the ground-state molecule because it has
(i) highly diagonal Franck-Condon factors,
(ii) a narrow linewidth,
and (iii) negligible leakage to other ({\it i.e.}, triplet) ground states.
To test our expectations, we experimentally investigated the rovibrational spectrum of ultracold $^{41}$K$^{87}$Rb molecules by using STIRAP and RF spectroscopy. From the experimental data, we determined the natural linewidth of $\mathrm{b^3\Pi_{0^+}} (v'=0)$ to be $\Gamma=2\pi \times
4.9(4)$ kHz while the Franck-Condon factor for the $\mathrm{X^1\Sigma^+} (v''=0) \rightarrow \mathrm{b^3\Pi_{0^+}} (v'=0)$ transition is $0.9474(1)$. Based on these values, we conclude that $^{41}{\rm K}^{87}{\rm Rb}$ is an ideal system for sub-microkelvin three-dimensional laser cooling of molecules.
By combining this narrow-line cooling with photoassociation and STIRAP, cooling of polar molecules to the sub-microkelvin temperatures is possible in an all-optical manner \cite{aika10}.

\begin{figure}
\includegraphics[width=6cm]{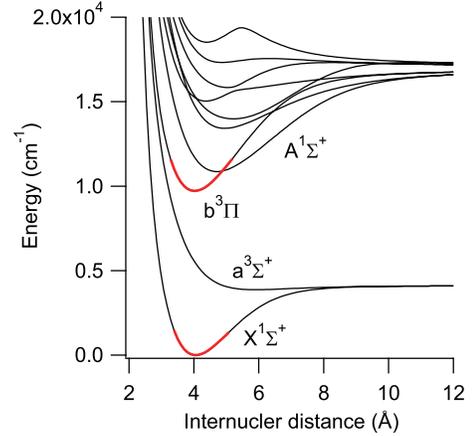}
\caption{\label{fig:1} (Color online)
Potential energy curves for KRb \cite{rous00}.
The curves for $\mathrm{X^1\Sigma^+}$ and $\mathrm{b^3\Pi}$ states are quite similar (highlighted in red), indicating highly diagonal Franck-Condon factors  for transitions between these two states. The decay of the $v'=0$ level of the $\mathrm{b^3\Pi_{0^+}}$ state into the $\mathrm{a^3\Sigma^+}$ state is strongly suppressed because of the extremely small transition dipole moments \cite{duli13private}.
}
\end{figure}

Figure 1 shows the potential energy curve of KRb as a function of internuclear separation.
The potential energy curves of $\mathrm{X^1\Sigma^+}$ and $\mathrm{b^3\Pi}$ look surprisingly similar, which signifies highly diagonal Franck-Condon factors for $\mathrm{X^1\Sigma^+} \rightarrow \mathrm{b^3\Pi_{0^+}}$ transitions.
Furthermore, radiative decay from $\mathrm{b^3\Pi_{0^+}} (v'=0)$
into the dissociative continuum of the $\mathrm{a^3\Sigma^+}$ triplet ground state
is negligibly small because of the extremely small electronic transition dipole moment of the $\mathrm{a^3\Sigma^+}$-$\mathrm{b^3\Pi_{0^+}}$ transition for short internuclear distances ($\sim 0.006\, e a_0$, where $a_0$ is the Bohr radius) \cite{duli13private,beuc06}.
The calculated decay rate of the $v'=0$ level of the $\mathrm{b^3\Pi_{0^+}}$ to $\mathrm{a^3\Sigma^+}$ is of the order of 1 Hz, which is negligibly small compared with the expected linewidth of the $\mathrm{b^3\Pi_{0^+}} (v'=0)$ state (a few kHz \cite{koto09}).

To test our prediction, we performed an experiment involving ultracold KRb molecules. Our experimental procedure for producing KRb molecules in the rovibrational ground state is described in detail in Ref. \cite{aika10}. In brief, KRb molecules in the X$^1\Sigma^+ (v=91, J=0)$ state (or $|\mathrm{X, 91, 0}\rangle$ state, for short) were made by photoassociation in a magneto-optical trap (MOT) of $^{87}$Rb and $^{41}$K atoms. Next, the molecules in $|\mathrm{X, 91, 0\rangle}$ were transferred to the $|\mathrm{X, 0, 0}\rangle$ state by STIRAP via the intermediate state $|\mathrm{(3)^1\Sigma^+, 41, 1}\rangle$ by using two lasers of wavelength 875 and 641 nm. 
By using 5 ns pulses from a tunable dye laser and a micro-channel plate, molecules were selectively observed based in their vibrational state by resonance-enhanced multi-photon ionization (REMPI) 
\begin{figure}
\includegraphics[width=8cm]{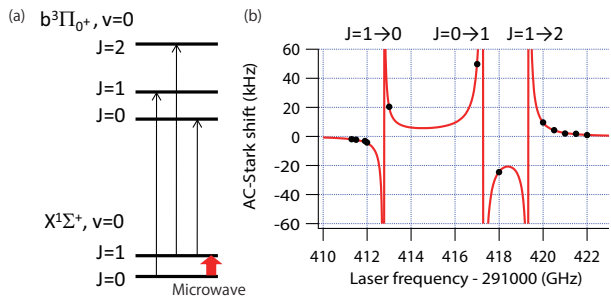}
\caption{\label{fig:search} (Color online) (a) Schematic of double-resonance spectroscopy for locating $|\mathrm{X,0}\rangle$-$|\mathrm{b,0}\rangle$ transition. The Stark shift induced by a laser beam close to the $|\mathrm{X,0}\rangle$-$|\mathrm{b,0}\rangle$ transition was detected by the microwave transition between the rotational levels in the $|\mathrm{X,0}\rangle$ state. (b) Observed Stark shifts. The measured Stark shifts showed a dispersive feature characteristic of ac-Stark shifts. The solid curve is a fit to the data for extracting resonance frequencies.}
\end{figure}

To date, the X$^1\Sigma^+ (v''=0) \rightarrow \mathrm{b^3\Pi_{0^+}} (v'=0) $ transition (or  $|\mathrm{X, 0}\rangle$-$|\mathrm{b, 0}\rangle$ transition, for short) of KRb has not been observed. The transition wavelengths predicted by {\it ab initio} calculations range from 1026.4 to 1036.9 nm \cite{yian96,park00,rous00,koto09}, which is 9 orders of magnitude larger than the expected natural linewidth.
To overcome this uncertainty, we looked for a dispersive signal rather than an absorptive signal because the signal falls off as $1/\Delta$ instead of $1/\Delta^2$ (where $\Delta$ is the detuning from resonance).
 We monitored the microwave transition frequency between rotational levels ($|\mathrm{X, 0, 0}\rangle$ and $|\mathrm{X, 0, 1}\rangle$), and looked for the ac-Stark shift induced by near-resonant light (Fig. \ref{fig:search}(a)).
  By using this method, we obsereved 3 transitions $|\mathrm{X, 0}\rangle$-$|\mathrm{b, 0}\rangle$, $|\mathrm{X, 0}\rangle$-$|\mathrm{b, 1}\rangle$, and $|\mathrm{X, 0}\rangle$-$|\mathrm{b^3\Pi_1, 0}\rangle$ with clearly resolved rotational structures (Fig. \ref{fig:search}(b)). Their transition wavelengths without rotational energies are 1028.7397(7), 1020.9746(7), and 1022.6870(7) nm, respectively, where the accuracies were limited only by the accuracy of our wavemeter.

\begin{figure}
\includegraphics[width=6cm]{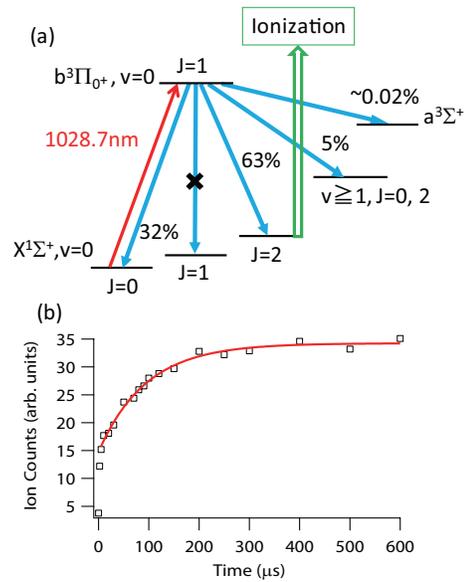}
\caption{\label{fig:pumping} (Color online) (a) Schematic of optical pumping spectroscopy of $|\mathrm{b, 0, 1}\rangle$ state (see text). The branching ratios from $|\mathrm{b, 0, 1}\rangle$ were calculated using the H\"{o}nl-London and Franck-Condon factors.
Most of the molecules in the $|\mathrm{X, 0, 0}\rangle$ state were optically pumped into the $|\mathrm{X, 0, 2}\rangle$ state by the 1028.7 nm laser which was resonant with the transition between $|\mathrm{X, 0, 0}\rangle$ and $|\mathrm{b, 0, 1}\rangle$. Subsequently, molecules in $|\mathrm{X, 0, 2}\rangle$ were selectively detected by resonance-enhanced multiphoton ionization (REMPI) based on their rotational state \cite{cleaning}.
(b) Measurement of time constant of optical pumping in saturated regime. Ion counts of the molecules in $|\mathrm{X, 0, 2}\rangle$ were measured as a function of the pulse width of the optical pumping laser. The measured time constant for optical pumping (95(7) $\mu$s) was converted into the lifetime of $|\mathrm{b, 0, 1}\rangle$ (32.5(24) $\mu$s) by using branching ratios shown in panel (a).
}
\end{figure}

We determined the natural linewidth and Doppler broadening of the $|\mathrm{X, 0}\rangle$-$|\mathrm{b, 0}\rangle$ transition by directly exciting molecules with a narrow-linewidth laser \cite{aika11}.
Figure \ref{fig:pumping}(a) shows a schematic of the experiment. Because of the almost diagonal Franck-Condon factors, the excitation optically pumped molecules into different rotational states (in this case, mostly to $|\mathrm{X, 0, 2}\rangle$).
The number of molecules in $|\mathrm{X, 0, 2}\rangle$ was detected by REMPI, which is sensitive to the rotational state \cite{cleaning}.
First, we saturated the transition by using a laser beam with an intensity 3 orders of magnitude greater than the saturation intensity.
Here, power broadening of the transition was larger than the hyperfine splitting in the ground state, which was $\sim 2\pi\times5$ kHz for $|\mathrm{X, 0, 0}\rangle$ \cite{alde08,brow03}.
From the optical pumping time constant (Fig. \ref{fig:pumping}(b)) and the branching ratios (Fig. \ref{fig:pumping}(a)),
we obtained the lifetime and natural linewidth of $|\mathrm{b,0,1}\rangle$ state; 32.5(24) $\mu$s and $2\pi\times$ 4.9(4) kHz, respectively.
Second, we measured the molecule temperature by observing the optical pumping spectrum in the nonsaturating regime. The typical temperature and mean velocity were 130 $\mu$K and 130 mm/s, respectively. This velocity is about 43 times the recoil velocity of the $|\mathrm{X,0}\rangle$-$|\mathrm{b,0}\rangle$ transition.

\begin{figure}[b]
\includegraphics[width=7cm]{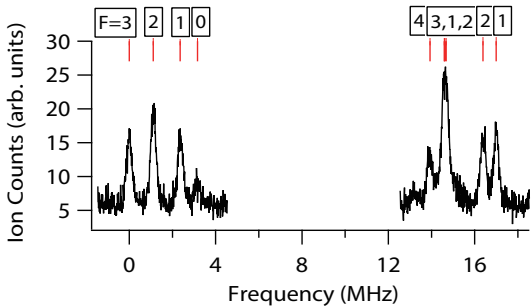}
\caption{\label{fig:bstate} (Color online) Spectrum of $|\mathrm{b, 0, 1}\rangle$ state obtained by optical pumping as described in Fig. \ref{fig:pumping}(a). The horizontal axis shows the frequency of the optical pumping laser relative to the left most signal. The hyperfine structure of $|\mathrm{b, 0, 1}\rangle$ was observed.}
\end{figure}

Franck-Condon factors are sensitive to rotational constants because both are directly related to the equilibrium internuclear distance. To exploit this, we performed detailed analysis of $|\mathrm{b, 0, 0}\rangle$, $|\mathrm{b, 0, 1}\rangle$, $|\mathrm{b, 1, 0}\rangle$, and $|\mathrm{b, 1, 1}\rangle$ states by acquiring hyperfine-resolved optical pumping spectra (Fig.\ \ref{fig:bstate}).
We assigned the hyperfine structures based on the Hamiltonian discussed in Refs. \cite{alde08,brow03}, which describes the interactions between two atomic nuclear spins and the rotation of diatomic molecules \cite{hyper}. The potential energy curve for a ground-state molecule was also refined by using STIRAP spectra of $|\mathrm{X, 1, 0}\rangle$ and $|\mathrm{X, 1, 2}\rangle$ states.

\begin{table}[t]
\begin{center}
\caption{\label{tbl:1} Experimentally obtained rotational and vibrational energies for molecular potentials of $\mathrm{X^1\Sigma^+}$ and $\mathrm{b^3\Pi_{0^+}}$. $E(v=1) - E(v=0)$ is the energy difference between $v=0 (J=0)$ and $v=1 (J=0)$ states. The accuracy was  limited by the accuracy of our wave meter. $B$ is the rotational constant whose accuracy is determined by the signal-to-noise ratio of the spectrum.
}
\begin{tabular}{c c c}\hline\hline
 & X$^1\Sigma^+$ & b$^3\Pi_{0^+}$ \\ \hline
$E(v=1)-E(v=0)$ (cm$^{-1}$)& 73.845(3) &73.936(3)\\ \hline
$B(v=0)$ (MHz)& 1095.3772(1) & 1118.37(1)\\ \hline
$B(v=1)$ (MHz)& 1091.93(1) & 1116.26(1) \\ \hline\hline
\end{tabular}
\end{center}
\end{table}

\begin{table}[t]
\begin{center}
\caption{\label{tbl:2} Obtained parameters for molecular potentials of $\mathrm{X^1\Sigma^+}$ and $\mathrm{b^3\Pi_{0^+}}$. Similarity between two potentials can be seen in values of $a_2, a_3$, and $r_0$ (see text).
}
\begin{tabular}{c c c}\hline\hline
 & X$^1\Sigma^+$ & b$^3\Pi_{0^+}$ \\ \hline
$a_2$ (cm$^{-1}$ \AA $^{-2}$)& 2286.3(2) &2279.2(2) \\ \hline
$a_3$ (cm$^{-1}$ \AA $^{-3}$)& -1133.5(8)  &-906.5(1.5)  \\ \hline
$r_0$ (\AA)& 4.067792(5) &4.02698(2) \\ \hline
$V_0$ (cm$^{-1}$)& 0 & 9720.531(3) \\ \hline\hline
\end{tabular}
\end{center}
\end{table}

\begin{figure}[b]
\includegraphics[width=5cm]{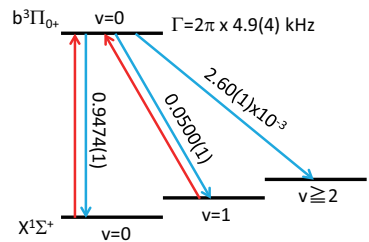}
\caption{\label{fig:FCF} (Color online) Franck-Condon factors and natural linewidth for $\mathrm{X^1\Sigma^+}$-$\mathrm{b^3\Pi_{0^+}}$ transitions. A cooling laser $|\mathrm{X,0,1}\rangle$-$|\mathrm{b,0,0}\rangle$ and repumping laser $|\mathrm{X,1,1}\rangle$-$|\mathrm{b,0,0}\rangle$ can form a highly closed cooling transition.}
\end{figure}

 From the spectra, we determined vibrational and rotational energies (Table \ref{tbl:1}). The X-state values agree well with the predictions of Ref. \cite{pash07}. To calculate the Franck-Condon factors, we determined the potential energy curves of X and b states. We assumed that the potential energy curves in the low energy region of X and b states could be written as
\begin{equation}
V(r)=\sum_{n=2}^{n_{max}}a_n(r-r_0)^n +V_0
\end{equation}
where $r$ is the internuclear distance.
Setting $n_{max}=3$ and fitting these curves to the experimental data, we determined $a_2, a_3, r_0$ and $V_0$ for X and b states (see Table \ref{tbl:2}) \cite{r0}. Because the difference in equilibrium distance between X and b states ($\sim$4.1 pm) is much smaller than the harmonic oscillator lengths for both potentials ($\sim$80 pm), the Franck-Condon factors between these states are highly diagonal. Using the obtained potential energy curves, we determined the Franck-Condon factors of$|\mathrm{X, 0}\rangle$-$|\mathrm{b,0}\rangle$, $|\mathrm{X, 1}\rangle$-$|\mathrm{b,0}\rangle$, and $|\mathrm{X, v \ge 2}\rangle$-$|\mathrm{b,0}\rangle$ transitions to be 0.9474(1), 0.0500(1), and $2.60(1)\times 10^{-3}$, respectively
(Fig. \ref{fig:FCF}).

We now discuss using the $\mathrm{X^1\Sigma^+} \rightarrow \mathrm{b^3\Pi_{0^+}}$ transition for realizing narrow-line cooling of KRb molecules. The proposed scheme is as follows; Ground-state KRb molecules at a few hundred microkelvin are prepared by photoassociation in a dual-species MOT followed by STIRAP. The cooling beam is red-detuned from the  $|\mathrm{X,0,1}\rangle$-$|\mathrm{b,0,0}\rangle$ transition, whereas the repumping beam is resonant with the $|\mathrm{X,1,1}\rangle$-$|\mathrm{b,0,0}\rangle$ transition.
Both beams irradiate molecules from all 6 directions.
Because the cooling beam is red detuned, molecules should predominantly scatter photons from the counter-propagating beam, and therefore be slowed down. 
Actual spectra of the cooling and repumping beams have to be optimized both for accessing all the hyperfine states and increasing the ranges of the capture velocity.

Several necessary conditions must be satisfied for cooling the molecules.
First, the cooling transition should be highly closed.
From the experimentally obtained Franck-Condon factor, KRb molecules can scatter $\sim 1/2.60(1) \times 10^{-3}$ $\sim 400$ photons before being pumped into other vibrational levels.
This number is sufficiently large for capturing molecules in this setup because the mean velocity of the initial cloud was $\sim 43$ times the recoil velocity.

 Second, the dark Zeeman substates, which are inevitably formed in this ($J''=1 \rightarrow J'=0$) type of rotationally closed transition, should be eliminated. As described in Ref. \cite{shum09}, these dark states can be mixed with bright states by applying a proper magnetic field such that the Larmor precession frequency in the ground state nearly equals the natural linewidth of the cooling transition. For the $^{41}$K$^{87}$Rb molecule, this condition can be fulfilled by a magnetic field of about 4 G, which is easy to use in the experiment.

 Finally, the cooling transition should be sufficiently strong. The natural linewidth obtained for the $\mathrm{X^1\Sigma^+} \rightarrow \mathrm{b^3\Pi_{0^+}}$ transition is $2\pi\times4.9(4)$ kHz; therefore the thermal expansion of the molecular cloud during the laser cooling ($\sim$ 6 ms \cite{coolingtime}) is estimated to be only $\sim 1$ mm. In addition, the maximum acceleration for this transition, which is $\sim 22$ m/s$^2$ (=$\hbar k \Gamma/4m$, where $m$ is the mass of the molecule), is more than twice the acceleration due to gravity.

 From the above considerations, we conclude that three-dimensional laser cooling can be achieved by this system.
Three-dimensional MOT is also a possible extension \cite{humm13}. Because the natural linewidth of the cooling transition is comparable to its recoil frequency ($2\pi\times1.5$ kHz), the laser cooling to the recoil temperature (140 nK) can be realized, as was successfully performed with strontium atoms \cite{kato99PRL}.

 In summary, we investigated the optical transitions between the low-lying vibrational states of $\mathrm{X^1\Sigma^+}$ and $\mathrm{b^3\Pi_{0^+}}$ states for KRb molecule and determined its natural linewidth and Franck-Condon factors. These results suggest a novel all-optical scheme for producing sub-microkelvin molecules, where molecules formed by photoassociation (and STIRAP) are cooled to the photon recoil temperature by three-dimensional laser cooling using $|\mathrm{X,0,1}\rangle$-$|\mathrm{b,0,0}\rangle$ and $|\mathrm{X,1,1}\rangle$-$|\mathrm{b,0,0}\rangle$ transitions. Because alkali-metal dimers have similar properties, this scheme is applicable to not only other isotopic KRb molecules but also other homo and heteronuclear alkali-metal-dimers. Because this scheme allows sub-microkelvin molecules to be rapidly produced, it could prove a powerful method for molecular precision spectroscopies. Furthermore, optically trapping molecules cooled by this scheme can be a novel strategy for realizing a quantum degenerate molecular gas.

We thank Olivier Dulieu for providing his calculations of transition dipole moments of KRb.
This work was supported by a Grant-in-Aid for Young Scientists (A) of JSPS (No. 23684034) and a Grant-in-Aid for Scientific Research on Innovative Areas of JSPS (No. 24104702).
K.A. acknowledges the support of the Japan Society for the Promotion of Science.

\bibliography{apssamp}

\begin{thebibliography}{99}
\bibitem{cleaning}For clarity, the molecules remaining in $|\mathrm{X, 0, 0}\rangle$ are optically pumped into higher vibrational states before ionization.
\bibitem{hyper} We will discuss the quantitative analyses of the hyperfine structures in detail elsewhere.
\bibitem{r0} The obtained equilibrium distance of the X state did not completely match the value in the Ref.\cite{pash07}.
This small difference should come from the models describing potential energy curves.
\bibitem{coolingtime} The time it takes for capturing molecules traveling with 43 times the recoil velocity can be estimated as
$t_{\rm capture}\sim \frac{1}{2 \pi \times 5 {\rm kHz}} \times 4 \times 43 \sim 6 \,{\rm ms}$.
\end{thebibliography}

\end{document}